\documentclass[twocolumn]{aastex631}

\usepackage{amsmath}
\usepackage{natbib}
\usepackage{multirow}
\usepackage{array}
\usepackage{tabularx}
\usepackage{booktabs}
\usepackage{graphicx}
\newcolumntype{P}[1]{>{\centering\arraybackslash}p{#1}}
\shorttitle{Kilonova-like emission from merger of NS-WD}
\shortauthors{Liu et al.} 

\begin{document}
\title{Neutron Star-White Dwarf Merger as One Possible Optional Source of Kilonova-like Emission: Implications for GRB 211211A}

\author[0009-0003-6940-2171]{Xiao-Xuan Liu} 
\affiliation{Guangxi Key Laboratory for Relativistic Astrophysics, Department of Physics, Guangxi University, Nanning 530004, China}

\author[0000-0001-6396-9386]{Hou-Jun L\"{u}} 
\altaffiliation{Corresponding author (LHJ) email: lhj@gxu.edu.cn}
\affiliation{Guangxi Key Laboratory for Relativistic Astrophysics, Department of Physics, Guangxi University, Nanning 530004, China}

\author[0009-0006-8625-5283]{Qiu-Hong Chen}
\affiliation{Guangxi Key Laboratory for Relativistic Astrophysics, Department of Physics, Guangxi University, Nanning 530004, China}

\author[0000-0002-4375-3737]{Zhao-Wei Du}
\affiliation{Guangxi Key Laboratory for Relativistic Astrophysics, Department of Physics, Guangxi University, Nanning 530004, China}

\author[0000-0002-7044-733X]{En-Wei Liang}
\affiliation{Guangxi Key Laboratory for Relativistic Astrophysics, Department of Physics, Guangxi University, Nanning 530004, China}
 
\begin{abstract}
Long-duration GRB 211211A, which lacked an associated supernova at very a low redshift $z=0.076$ but was associated with a possible kilonova emission, has attracted great attention. The neutron star-white dwarf (NS-WD) merger is proposed as a possible progenitor of GRB 211211A, and it could naturally explain the long duration of the prompt emission. However, the NS-WD merger is not an ideal site for producing heavy elements via r-process nucleosynthesis. In this Letter, we investigate the heavy elements produced in NS-WD mergers based on numerical simulations of nucleosynthesis via SkyNet and then calculate the resulting kilonova-like emission to compare with the solidly observed case of possible kilonova emission associated with GRB 211211A. By adopting three models (i.e., Model-A, Model-B, and Model-C) from M. A. R. Kaltenborn et al. at different temperatures ($T=4$ GK, 5 GK, and 6 GK), which are treated as free parameters, we find that the mass number of the heaviest element produced in our simulations is less than 90 ($A< 90$). Moreover, by comparing the calculated kilonova-like emission with the afterglow-subtracted observations of the possible kilonova associated with GRB 211211A, it is found that the merger of an NS and WD cannot be ruled out as the origin of GRB 211211A to produce the possible kilonova emission if the remnant of the WD-NS merger is a supramassive or stable magnetar. Otherwise, it is difficult to explain the early possible kilonova emission following GRB 211211A by invoking the merger of a WD and an NS.

\end{abstract}

\section{Introduction} \label{sec:intro}
Gamma-ray bursts (GRBs), the most luminous electromagnetic explosions in the Universe, are traditionally classified into two categories based on their duration: short-duration GRBs (SGRBs; $T_{90}<2s$) and long-duration GRBs (LGRBs; $T_{90}>2s$) \citep{1993ApJ...413L.101K}. Robust associations of the underlying supernovae (SNe) with some LGRBs suggest that they are likely related to the collapse of massive stars \citep{1993ApJ...405..273W, 1998Natur.395..670G, 2003ApJ...591L..17S, 2006Natur.441..463F, 2006Natur.442.1011P, 2006ARA&A..44..507W}. In contrast, some SGRBs are associated with nearby host galaxies with little star formation, and others do not have an associated  SN \citep{2005Natur.438..988B, 2005Natur.437..845F, 2005Natur.437..859H, 2005Natur.438..991T}, suggesting that the SGRBs are possibly originated from the merger of binary compact stars, such as binary neutron star (BNS) mergers \citep{1986ApJ...308L..43P, 1989Natur.340..126E} and neutron star-black hole (NS-BH) mergers \citep{1991AcA....41..257P}. On the other hand, such binary systems are also considered to be potential sources of gravitational-wave radiation and kilonova emission \citep{2014ARA&A..52...43B, 2018pgrb.book.....Z}. The first ``smoking gun'' evidence to confirm the BNS merger as the source of SGRBs, gravitational-wave radiation, and kilonova, emerged from the direct detection of GW 170817 via Advanced LIGO and Virgo, which was associated with electromagnetic counterparts, such as GRB 170817A and AT 2017gfo \citep{2017PhRvL.119p1101A, 2017ApJ...848L..21A, 2017Natur.551...64A, 2017ApJ...848L..19C, 2017Sci...358.1556C, 2017NatAs...1..791C, 2017Sci...358.1570D, 2017Sci...358.1565E, 2017ApJ...848L..14G, 2017Natur.551...80K, 2017Sci...358.1559K, 2017Natur.551...67P, 2017ApJ...848L..15S, 2017Sci...358.1574S, 2017Natur.551...75S, 2017PASJ...69..102T, 2017ApJ...848L..27T, 2018NatCo...9..447Z}.

Except for the confirmed event  GW 170817/ GRB 170817A/ AT 2017gfo, some SGRBs have been associated with kilonova candidates, such as GRB 050709 \citep{2016NatCo...712898J}, GRB 130603B \citep{2013ApJ...774L..23B, 2013Natur.500..547T} and GRB  160821B \citep{2017ApJ...835..181L, 2019ApJ...883...48L, 2019MNRAS.489.2104T}. However, from an observational point of view, it is not only SGRBs that exhibit the features of the merger but also several LGRBs, which are more likely SGRBs with extended emission (EE-SGRB), showing potential kilonova signatures but lacking an associated SN, for example, GRB 060614 \citep{2006Natur.444.1050D, 2006Natur.444.1047F, 2006Natur.444.1053G, 2006Natur.444.1044G, 2008MNRAS.385.1455M, 2015NatCo...6.7323Y}, GRB 211211A \citep{2022Natur.612..223R, 2022Natur.612..228T, 2022Natur.612..232Y, 2023ApJ...943..146C, 2023NatAs...7...67G}, GRB 211227A \citep{2022ApJ...931L..23L, 2023A&A...678A.142F}, and GRB 230307A \citep{2023arXiv230800633G, 2024ApJ...962L..27D, 2024Natur.626..737L, 2024Natur.626..742Y, 2025NSRev..12E.401S}. On the other hand, the short GRB 200826A, which seems to be associated with an SN, suggests an origin from the collapse of massive stars \citep{2021NatAs...5..917A, 2021NatAs...5..911Z}. In summary, the traditional long-versus-short-GRB scheme no longer always corresponds to the two distinct physical origins \citep{2006Natur.444.1010Z, 2010ApJ...725.1965L, 2013ApJ...774L..23B, 2014MNRAS.442.1922L}.

One interesting question focuses on the EE-SGRBs; it is challenging for the BNS mergers and NS-BH mergers to explain the duration of their prompt $\gamma$-ray emission due to their shorter timescales of the mergers. Thus, the neutron star-white dwarf (NS-WD) merger has been proposed to explain the observational properties of EE-SGRBs \citep{2022Natur.612..232Y, 2023ApJ...947L..21Z, 2024ApJ...964L...9W}. The reason is that WDs are less dense than NSs, and the free-fall timescale of NS-WD mergers could be significantly longer than that of BNS mergers or NS-BH mergers. If this is the case, it could naturally explain the long timescale characteristics of these EE-SGRBs. However, WDs are proton rich, so the ejecta from the NS-WD mergers may not be as neutron rich as those from systems like NS-NS mergers, NS-BH mergers, collapsars, etc. Therefore, the astrophysical environment of the NS-WD mergers seems not to be suitable for r-process element production \citep{2019MNRAS.488..259F, 2022MNRAS.510.3758B, 2023ApJ...956...71K}.

To understand the process of NS-WD mergers, we provide a relatively clear picture of NS-WD mergers: NS-WD mergers are originated from binaries formed through common envelope evolution and gravitational-wave-driven orbital decay. Population synthesis suggests that such systems are abundant in the Galaxy \citep{2001A&A...365..491N}. When unstable mass transfer begins, the WD is tidally disrupted to form an accretion disk around the NS \citep{1984MNRAS.208..721P, 1999ApJ...520..650F}. Numerical simulations have extensively explored the subsequent disk evolution, which involves nuclear burning, fallback accretion, angular momentum transport, and outflows driven by neutrino cooling or magnetic processes. These features are somewhat similar to that of found in postmerger disks from NS-NS systems \citep{2013MNRAS.435..502F}. Recent works have further investigated how the structure, composition, and thermodynamic state of the disk affect the ejecta mass, nucleosynthesis, and the properties of the resulting faint and rapidly evolving transients \citep{2019MNRAS.486.1805Z, 2020MNRAS.493.3956Z, 2022MNRAS.510.3758B, 2023ApJ...956...71K}. Especially, \cite{2024MNRAS.527.5540C} provide insight into the radioactive heating of heavy elements from nuclear burning, which may inform further studies of the NS-WD merger ejecta.

It is interesting to ask the following two questions: What is the approximate mass number of the heaviest element that can be produced by NS-WD mergers? How bright is the kilonova-like emission corresponding to it?

In this work, we investigate the heavier elements produced by NS-WD mergers based on nucleosynthesis simulations. Then, we explore the luminosity of the corresponding kilonova-like emission produced by the radioactive decay of the elements produced by nucleosynthesis based on detailed numerical calculations. Finally, we compare the calculated luminosity of the kilonova-like emission with the observations of the possible kilonova associated with GRB 211211A. This Letter is organized as follows. In Section \ref{sec:method}, we introduce the details of nucleosynthesis and the method we adopted to calculate the kilonova-like light curves. The results of the nucleosynthesis elements and the corresponding kilonova-like light curves are presented in Section \ref{sec:result}. The application to a possible kilonova emission associated with GRB 211211A based on our results is shown in Section \ref{sec:211211A}. Conclusions are drawn in Section \ref{sec:end}, along with some additional discussion. Throughout this Letter, we adopt a concordance cosmology with the parameters $H_{\rm 0}=71~\rm km~s^{-1}~Mpc^{-1}$, $\Omega_{\rm M}=0.30$, and $\Omega_{\Lambda}=0.70$.

\section{Methods}\label{sec:method}
In general, based on the mass of white dwarfs ($M_{\rm WD}$) in WD-NS binaries, there are two main evolutionary pathways that are widely discussed for such binary systems, namely stable mass transfer ($M_{\rm WD}<M_{\rm WD,c}$) and unstable mass transfer ($M_{\rm WD}>M_{\rm WD,c}$), where $M_{\rm WD,c}\sim 0.2 M_\odot$ is the critical mass of the WD \citep{2016MNRAS.461.1154M}. More than 20 WD-NS binaries have been discovered in our Galaxy, and the mass of WDs in such binaries is larger than the critical mass \citep{2010ApJ...715..230O}. It suggests that the majority of WD-NS binary systems will undergo a merger with the process of unstable mass transfer within a Hubble time \citep{2018A&A...619A..53T}. In our work, we adopt the unstable mass transfer model and select three different groups of WD-NS systems for simulation, with mass groups of (0.75-1.4) $M_\odot$, (1-1.25) $M_\odot$, and (1-1.8) $M_\odot$, which we named as Model-A, Model-B, and Model-C, respectively (see Table \ref{tab:1}).

\begin{table*}[htbp]
    \centering 
    \setlength{\tabcolsep}{4pt}
    \begin{tabular*}{\linewidth}{@{\extracolsep{\fill}}ccccccc@{}}
        \toprule
          & $M_{\rm WD}$ ($M_\odot$) & $M_{\rm NS}$ ($M_\odot$) & q & $\rho_{\rm disk}$ (g $\rm cm^{-3}$) & $\rho_{\rm ej}$ (g $\rm cm^{-3}$) & $M_{\rm ej}$ ($M_\odot$) \\
        \midrule
        Model-A & 0.75 & 1.4 & 0.53 & 3.64e4 & 3.00e5 & 0.053\\
        Model-B & 1 & 1.25 & 0.80 & 1.35e5 & 4.42e5 & 0.078\\
        Model-C & 1 & 1.8 & 0.56 & 9.84e4 & 3.68e5 & 0.065\\
        \bottomrule
      \end{tabular*} 
      \caption{Selected parameters for our simulations, such as mass of WD ($M_{\rm WD}$) and NS ($M_{\rm NS}$), mass ratio ($q = M_{\rm WD}/M_{\rm NS}$), initial density of the disk ($\rho_{\rm disk}$), initial density of the ejecta ($\rho_{\rm ej}$), and the mass of the ejecta ($M_{\rm ej}$). The parameters are derived from \cite{2023ApJ...956...71K} and \cite{2019MNRAS.486.1805Z}} \label{tab:1}
\end{table*}

\subsection{Nucleosynthesis}\label{sec:methodnuc}
The merger of binary compact stars (e.g., NS-NS, NS-BH) or SN explosions is an important site for the nucleosynthesis of heavy elements. One interesting question is how heavy the heaviest element can be formed during the WD-NS mergers. In this section, we perform the nucleosynthesis via the nuclear reaction network code SkyNet to do the numerical simulations \citep{2015ApJ...815...82L, 2017ApJS..233...18L}.

SkyNet can compute the evolution of nucleosynthesis in a wide range of astrophysical scenarios where nucleosynthesis occurs, and it has been widely used to evolve the abundance of nuclear species under the influence of nuclear reactions. The network contains 7843 nuclides and 140,000 reactions, ranging from free neutrons ($Z=0$) and protons ($Z=1$) to $\rm {^{337}Cn}$ ($Z=112$). SkyNet adopts a modified version of the Helmholtz equation of state (EOS; \cite{2000ApJS..126..501T}) to calculate the entropy. The nuclear reaction rates used in SkyNet are taken from the JINA REACLIB database \citep{2010ApJS..189..240C}. Nuclear masses and partition functions are included in the WebNucleo XML file distributed with REACLIB for nuclide species that are without experimental data, and we use the theoretical data derived from the finite-range droplet model \citep{2015PhRvC..91b4310M}. 

The first step is to evolve the accretion disk, which is composed of material generated by the tidal disruption of the WD. The initial density profile of the disk that we adopt can be described in \cite{2016MNRAS.461.1154M}, and we adopt the average density as the initial condition density in our simulations. In terms of the temperature, we investigate the effect of different initial temperatures due to the poorly known temperature. Thus, three groups of temperatures, such as 4, 5, and 6 GK, are used in the simulations. We initialize the composition of the material in the disk, namely, equal parts of carbon and oxygen, and with 1\% helium. During the disk evolution, we assume that the overall density and temperature of the whole disk do not change too much, and we adopt a constant density and temperature trajectory in SkyNet to perform the simulations. The total timescale of disk evolution is set to 60 s, and one can obtain the final abundance of the disk evolution.

Then, we set the final abundance of the disk evolution as the initial abundance of the ejecta evolution. In our nucleosynthesis simulations of the ejecta, for the density profile, we use the parameterized trajectory provided by \cite{2015ApJ...815...82L}. The initial density of the ejecta is set to $\rho_{\rm ej}=3.00 \times 10^5 \rm~g\, cm^{-3}$ for Model-A, which is taken from the results of \cite{2019MNRAS.486.1805Z}. By assuming that the ejecta of Model-B and Model-C share the same radius as in Model-A, together with the results of numerical relativity simulations given by \cite{2023ApJ...956...71K}, one can derive the initial densities of Model-B and Model-C to be $4.42\, \times \, 10^5\, \rm g\, cm^{-3}$ and $3.68\, \times \, 10^5\, \rm g\, cm^{-3}$, respectively. Table \ref{tab:1} shows the details of the initial parameter settings for both disk and ejecta. Moreover, the expansion timescale is approximately set to be 3 s, considering the long duration of WD-NS mergers. For ejecta evolution, we also set three groups of initial temperatures, which are 4, 5, and 6 GK, respectively. Then, the temperature profiles are set as $TV^{1/3}=\rm const$ for the evolution of the ejecta, which is because the expansion of the ejecta is nearly adiabatic \citep{2018MNRAS.478.3298W, 2010ApJ...712.1359F}.

\subsection{Radioactive heating rate}\label{sec:MethodQdot}
A large number of elements can be produced through nucleosynthesis, releasing a substantial amount of energy through radioactive decay. The energy from the radioactive decay of these elements can heat ejecta through interactions. So, the total effective heating rate of the merger ejecta contains the sum of the contributions from all radioactive decay of elements, and it can be expressed as
\begin{equation}
    \dot{Q}(t) = f_{\rm tot}(t)\dot{q}(t).
\end{equation}
Here, $\dot{q}(t)$ is the radioactive heating rate for all nuclides, which can be directly calculated by SkyNet. $f_{\rm tot}(t)$ is the thermalization efficiency which is estimated by the analytic method proposed in \cite{2016ApJ...829..110B},
\begin{equation}
    f_{\rm tot}(t) = 0.36\bigg[{\rm exp}(-at) + \frac{{\rm ln}(1+2bt^d)}{2bt^d}\bigg],
\end{equation}
where $a$, $b$, and $d$ are constants. In our simulations and calculations, we fixed the values of $a=0.27$, $b=0.10$ and $d=0.60$ \citep{2016ApJ...829..110B}. 

\subsection{Kilonova-like emission}\label{sec:MethodKNe}
The kilonova emission, which is an optical/infrared transient, is generated in the ejected material and is powered by the radioactive decay of r-process nuclei \citep{1998ApJ...507L..59L,2010MNRAS.406.2650M,2015NatCo...6.7323Y,2018ApJ...852L...5M}. If extra energy is injected into such a transient, except for radioactive decay of r-process nuclei, the kilonova will be brighter and is called merger-nova \citep{2013ApJ...776L..40Y, 2017ApJ...837...50G, 2021ApJ...912...14Y}. In our work, the merger ejecta is assumed to be spherically symmetric \citep{2017Natur.551...80K}; even some numerical simulations of double NS merger suggest that the ejecta or radioactive decay of r-process is more or less anisotropic \citep{2017PhRvD..95f3016C,2018ApJ...858...52S,2023MNRAS.521.1858C}. By adopting an average velocity of ejecta $v_{\rm ej}=0.1$ c taken from the simulation result of \cite{2019MNRAS.486.1805Z}, and following the method in \cite{2024MNRAS.527.5540C}, we divide the merger ejecta into $n$ layers and treat the evolution of each layer independently. The velocities of the innermost layer (first layer) and the outermost layer (last layer) are set to $v_{\rm min}=v_{\rm ej}-0.03c$ and $v_{\rm max}=v_{\rm ej}+0.03c$, respectively. Thus, the expansion velocity of the $i$th layer is given by
\begin{equation}
    v_i=v_{\rm min}+(i-1)\frac{v_{\rm max}-v_{\rm min}}{n-1}.
\end{equation}
The radius of the $i$th layer at time $t$ is given by $R_{i}=v_{i}t$, and the mass of the $i$th layer is $m_i=\int_{R_i}^{R_{i+1}}{4\pi\rho(v_i, t) r^2dr}$. The density profile is assumed to be described as a broken power-law function, as adopted in \cite{2017Natur.551...80K}, namely, 
\begin{equation}
    \rho(v_i,t)=\begin{cases}
    \zeta_\rho\frac{M_{\rm ej}}{v_{\rm t}^3t^3}(\frac{v_i}{v_{\rm t}})^{-\delta}, \quad & v_i<v_{\rm t}, \\
     \zeta_\rho\frac{M_{\rm ej}}{v_{\rm t}^3t^3}(\frac{v_i}{v_{\rm t}})^{-n}, \quad & v_i\geq v_{\rm t}.
    \end{cases}
\end{equation}
Here, the default values for the exponents are $\delta=1$ and $n=10$. The transition velocity $v_{\rm t}$ between the inner and outer layers is given by
\begin{equation}
    v_{\rm t} = \zeta_{v}(\frac{2E}{M_{\rm ej}})^{1/2},
\end{equation}
where $E=\sum_im_iv_i^2/2$ is the sum of the kinetic energy of all layers, $\zeta_\rho$ and $\zeta_v$ are the normalization constants. Based on the mass conservation for a given $M_{\rm ej}=\int_{R_1}^{R_n}{4\pi\rho(v_i,t)r^2dr}$, $\zeta_\rho$ and $v_{\rm t}$ can be derived.

The main difference between the model we used and the toy model presented in \cite{2017LRR....20....3M, 2019LRR....23....1M} is that the rate of density decreasing for different mass layers in the ejecta is different. The densities of different mass layers in \cite{2019LRR....23....1M} are all decreasing at the same rate. However, in our model, we adopt the density profile used in \cite{2017Natur.551...80K}, i.e., the densities of different layers have different decreasing rates. The density of the inner layers decreases more slowly, while the density of the outer layers decreases faster. Within the scenario of the NS and WD merger, because of the continuous injection of material into the accretion disk (and the long duration of the process), the density profile we used should be more suitable.

Following the method in \cite{2019LRR....23....1M}, the evolution of thermal energy of each layer can be written as
\begin{equation}
    \frac{dE_i}{dt}=-\frac{E_i}{R_i}\frac{dR_i}{dt}-L_i+\dot{Q}(t)m_i,
\end{equation}
where $E_i$ is the internal energy of the $i$th layer and $\dot{Q}(t)$ is the heating rate, which can be given by Equation (1). $L_i$ is the radiation luminosity of each layer, which can be estimated as
\begin{equation}
    L_i=\frac{E_i}{t_{d,i}+t_{lc,i}},
\end{equation}
where $t_{d,i}=\tau_iv_it/c$ is the photon diffusion timescale and $t_{lc,i}=v_{i}t/c$ is the energy loss time limitation to the light-crossing time. The optical depth ($\tau_i$) of the $i$th shell is given by
\begin{equation}
    \tau_i=\int_{R_i}^\infty{\rho(v_i,t)\kappa dr},
\end{equation}
where $\kappa$ is the opacity of the ejecta. In our calculation, the ejecta are predominantly composed of iron-group elements; thus, we set $\kappa=0.2\rm cm^2 g^{-1}$, which is a reasonable approximation for iron-rich ejecta in \cite{2019LRR....23....1M}. The total bolometric luminosity is the sum of the contributions from all layers,
\begin{equation}
    L_{\rm bol}=\sum_i{L_i}.
\end{equation}
The effective temperature of the thermal emission is written as
\begin{equation}
    T_{\rm eff}=\Big(\frac{L_{\rm bol}}{4{\rm\pi}\sigma_{\rm SB}R_{\rm ph}^2}\Big)^{1/4},
\end{equation}
where $\sigma_{\rm SB}$ is the Stefan-Boltzmann constant and $R_{\rm ph}$ is the photospheric radius when the optical depth equals unity, i.e.,
\begin{equation}
    \tau_{\rm ph}=\int_{R_{\rm ph}}^\infty{\rho(v_i,t)\kappa dr=1}.
\end{equation}

For a given frequency $\nu$, the observed flux density could be calculated as
\begin{equation}
    F_\nu(t)=\frac{2\pi h\nu^3R_{\rm ph}^2}{c^2D_{L}^2}\frac{1}{{\rm exp}(h\nu/kT_{\rm eff})-1},
\end{equation}
where $h$ is Planck constant, $k$ is Boltzmann constant, and $D_{\rm L}$ is the luminosity distance of the source. Finally, the monochromatic AB magnitude is determined by
\begin{equation}
    M_\nu(t)=-2.5\log_{10}\big(\frac{F_\nu(t)}{3631\,{\rm Jy}}\big).
\end{equation}

\section{Results of numerical simulations}\label{sec:result}
\subsection{Numerical Simulations of Nucleosynthesis}\label{sec:ResultNuc}
\begin{figure}[!htpb]
    \centering
    \includegraphics[height=2.55in]{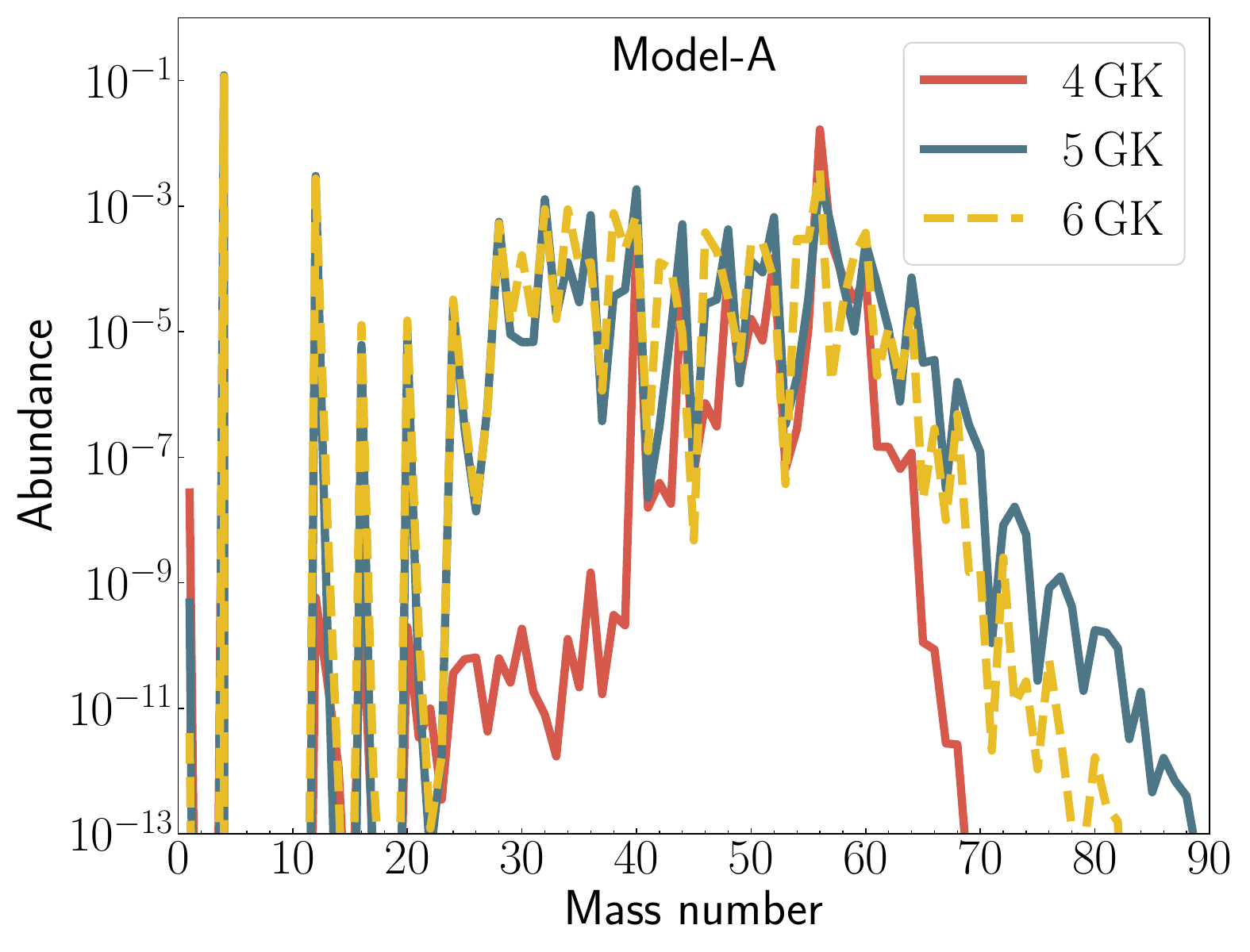}
    \hfill
     \includegraphics[height = 2.55in]{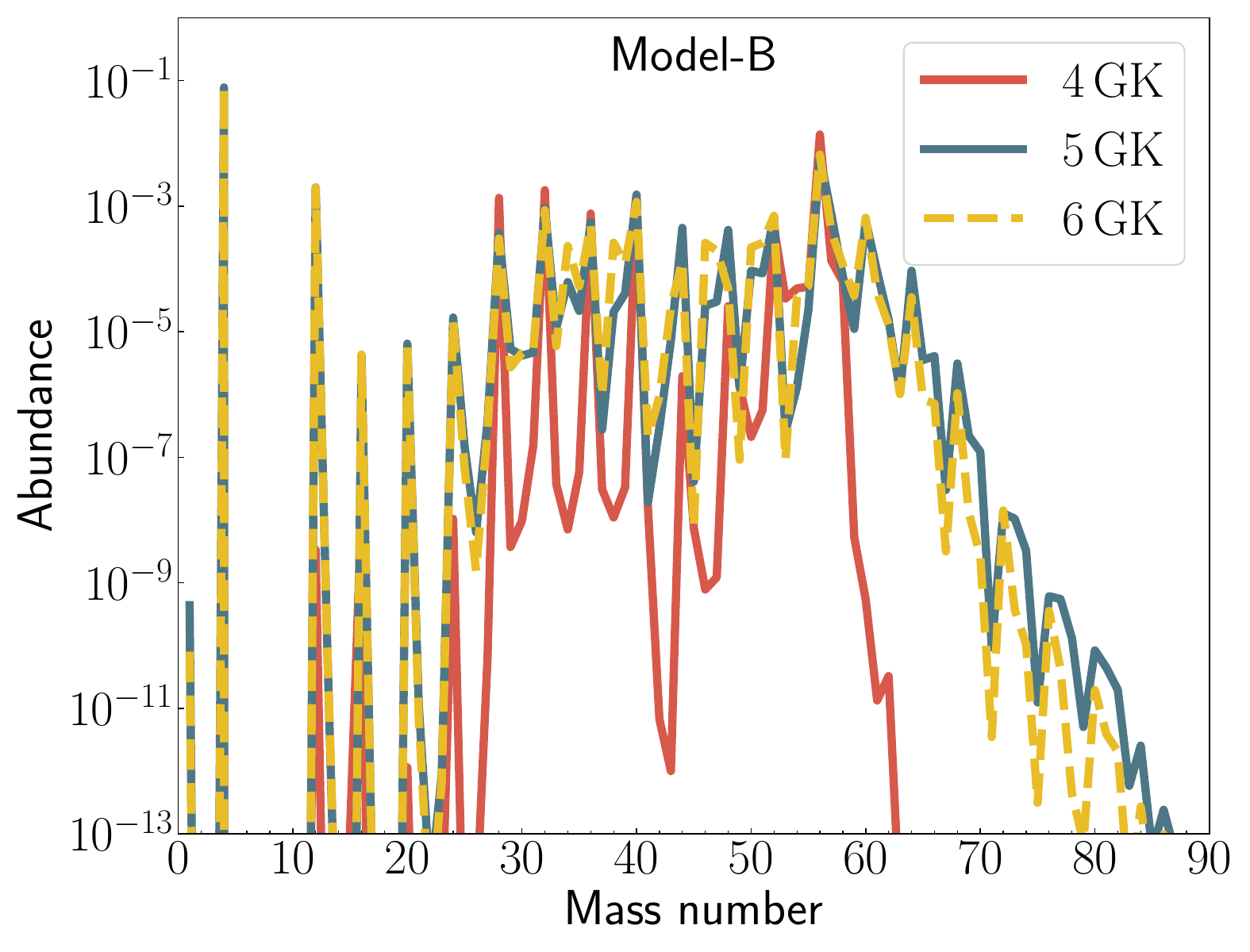}
     \hfill
      \includegraphics[height = 2.55in]{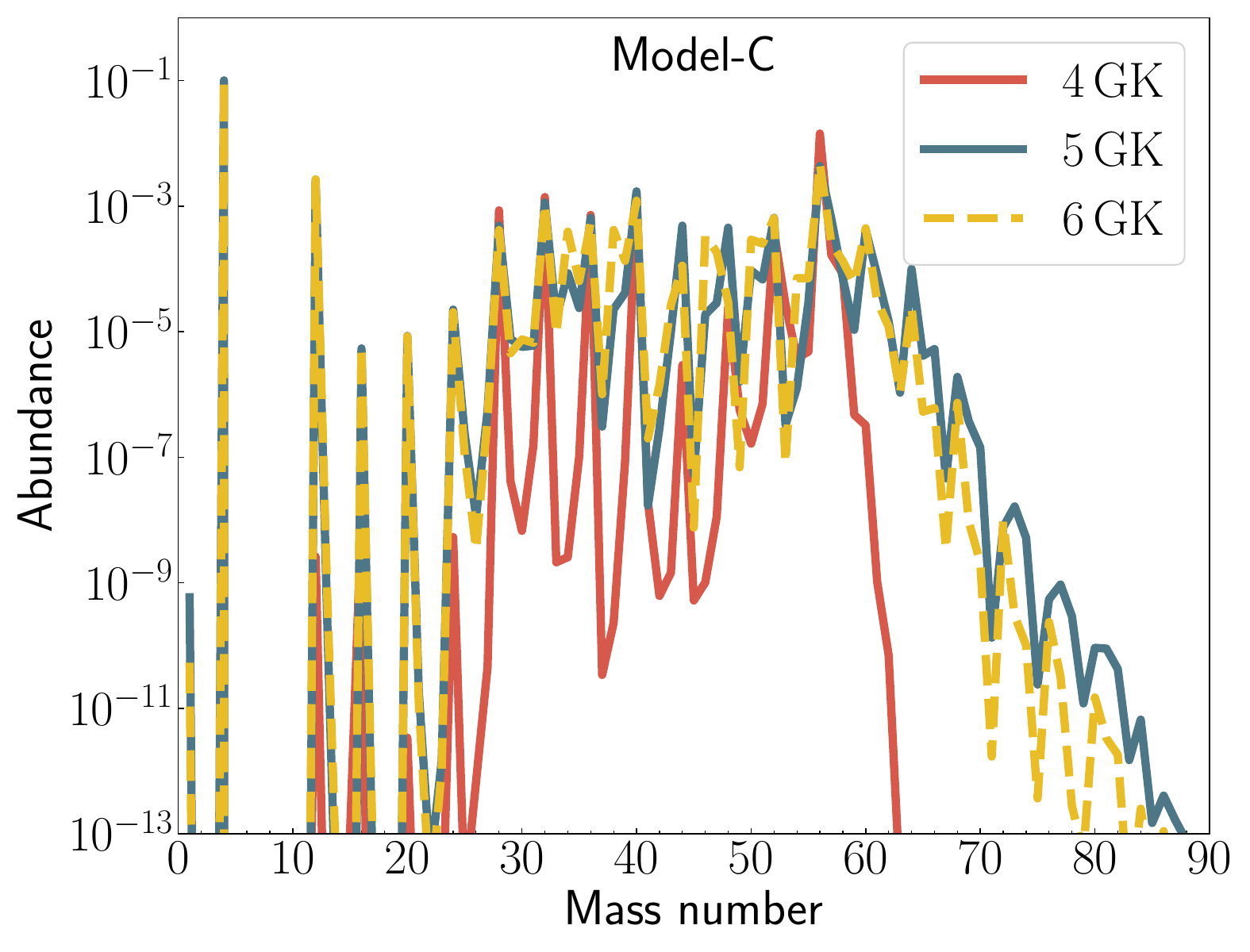}
    \caption{Final abundance of the ejecta for Model-A (top panel), Model-B (middle panel), and Model-C (bottom panel) with different temperatures ($T=4$, 5, and 6 GK), respectively.}
    \label{FIG:nuc}
\end{figure}

Based on the method and theories of nucleosynthesis mentioned in Section \ref{sec:methodnuc}, we perform numerical simulations of nucleosynthesis via SkyNet for both the disk and ejecta from WD-NS mergers. Figure \ref{FIG:nuc} shows the final abundance of the ejecta for three different groups of WD-NS systems with mass groups of (0.75-1.4) $M_\odot$ (Model-A), (1-1.25) $M_\odot$ (Model-B), and (1-1.8) $M_\odot$ (Model-C), respectively. Moreover, we also present the final abundance of the ejecta with different temperatures (i.e., 4, 5, and 6 GK) for each model. For Model-A, it is found that the final abundance of ejecta for temperature $T=5$ GK is not very different from that of temperature $T=6$ GK, and the nuclides produced by $T=5$ GK are only slightly heavier than those produced by $T=6$ GK. However, the abundance of heavier elements (i.e., heavier than iron) with $T=5$ GK and $T=6$ GK is greater than that produced by $T=4$ GK. By adopting different mass distributions of WD and NS in the binary, the results of Model-B and Model-C exhibit similar trends to those observed in Model-A.

On the other hand, at the given temperature of $T=4$ GK, we find that the abundances in all three models peak at mass number $A=56$, namely, the abundance of $\rm ^{56}Fe$ is the highest. If this is the case, the decay chain $\rm ^{56}Ni\rightarrow{}^{56}Co\rightarrow{}^{56}Fe$ can release a large amount of energy. In addition, the mass number of the heaviest elements it can produce is in the range of $60<A<70$. At the given temperatures of $T=5$ GK and $T=6$ GK, a significant amount of elements with mass number $A=56$ are still produced, but the major difference compared with the result at $T=4$ GK is that the abundance of $\rm ^{4}He$ is substantially higher. There are two primary ways to cause such a higher abundance of $\rm ^{4}He$. One is that the iron peak elements can be decomposed into free nuclei (protons, neutrons) and lighter nuclei (e.g., $\alpha$ particles) when the initial temperature is higher ($T\sim 5$ and 6 GK). With the ejecta rapidly expanding and cooling down, the nuclear reactions, which were initially in dynamic equilibrium, are "frozen", and the decomposed $\alpha$ particles are retained because they are too late to recombine into heavy nuclei. The other one is that much of $\rm ^{4}He$ can be produced by the combination of neutrons and protons at a higher temperature. With the ejecta expanding at high speed, the density of the ejecta decreases, which makes $\rm ^4He$ difficult to be consumed.

Moreover, by comparing the heavier elements of the three models at the same temperature, it is found that the mass number of the produced heavier nuclides follows $A_{\rm Model-A}>A_{\rm Model-C}>A_{\rm Model-B}$. This is because the entropy of these three models, which is related to the density as $s\propto{\rho}^{-1}$, is calculated using the Helmholtz EOS. As shown in Table \ref{tab:1}, the density of the ejecta in Model-A is the smallest among the three models, but it has the highest entropy. Therefore, Model-A tends to produce elements with higher mass numbers compared to the other two models. Based on Figure \ref{FIG:nuc}, no matter which temperature or model we choose , the mass number of the heaviest produced nuclides is less than 90 (i.e., $A<90$), and a large fraction of the produced elements are around the iron peak.

\begin{figure}[!htpb]
    \centering
   \includegraphics[height = 2.55in]{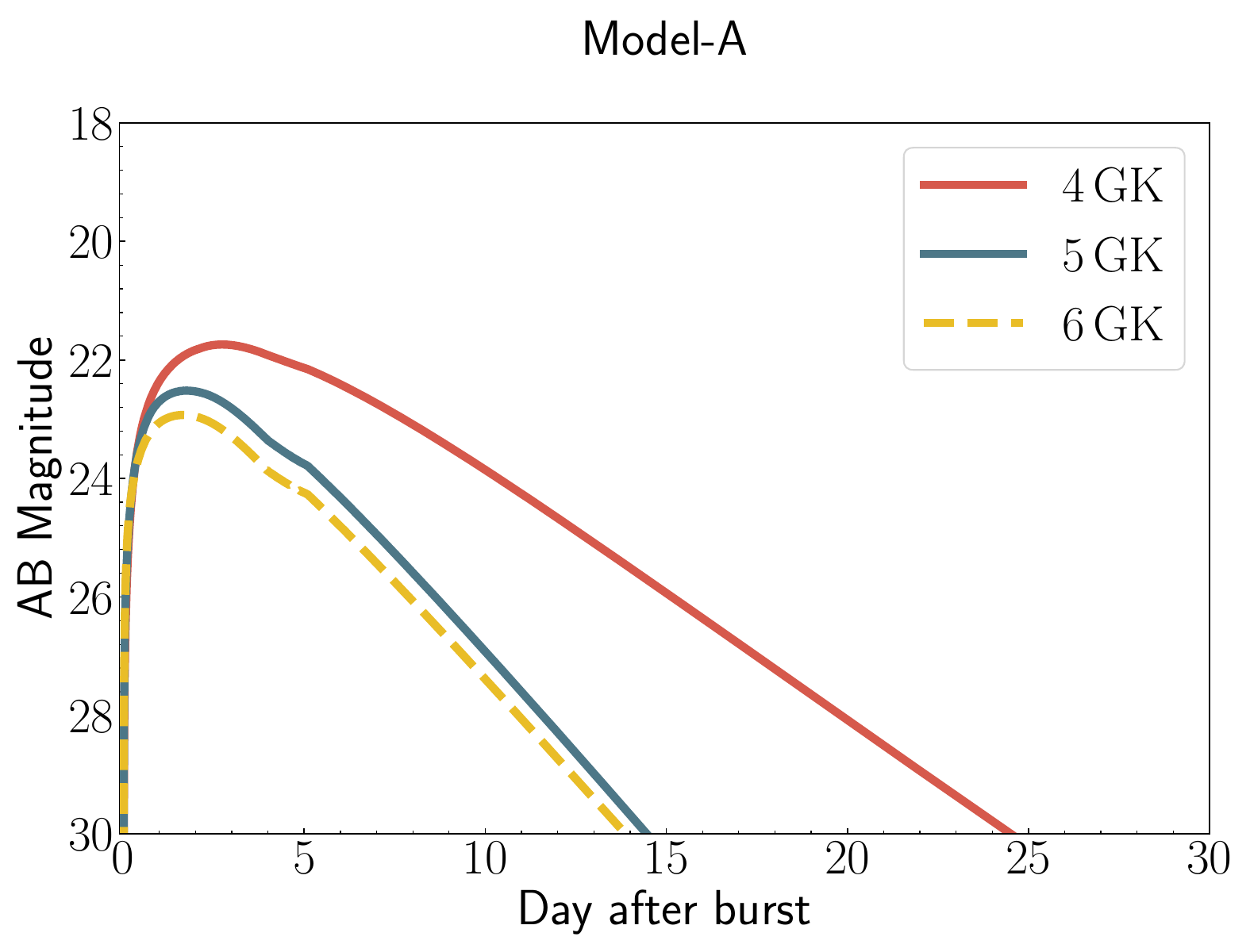}
   \hfill
   \includegraphics[height = 2.55in]{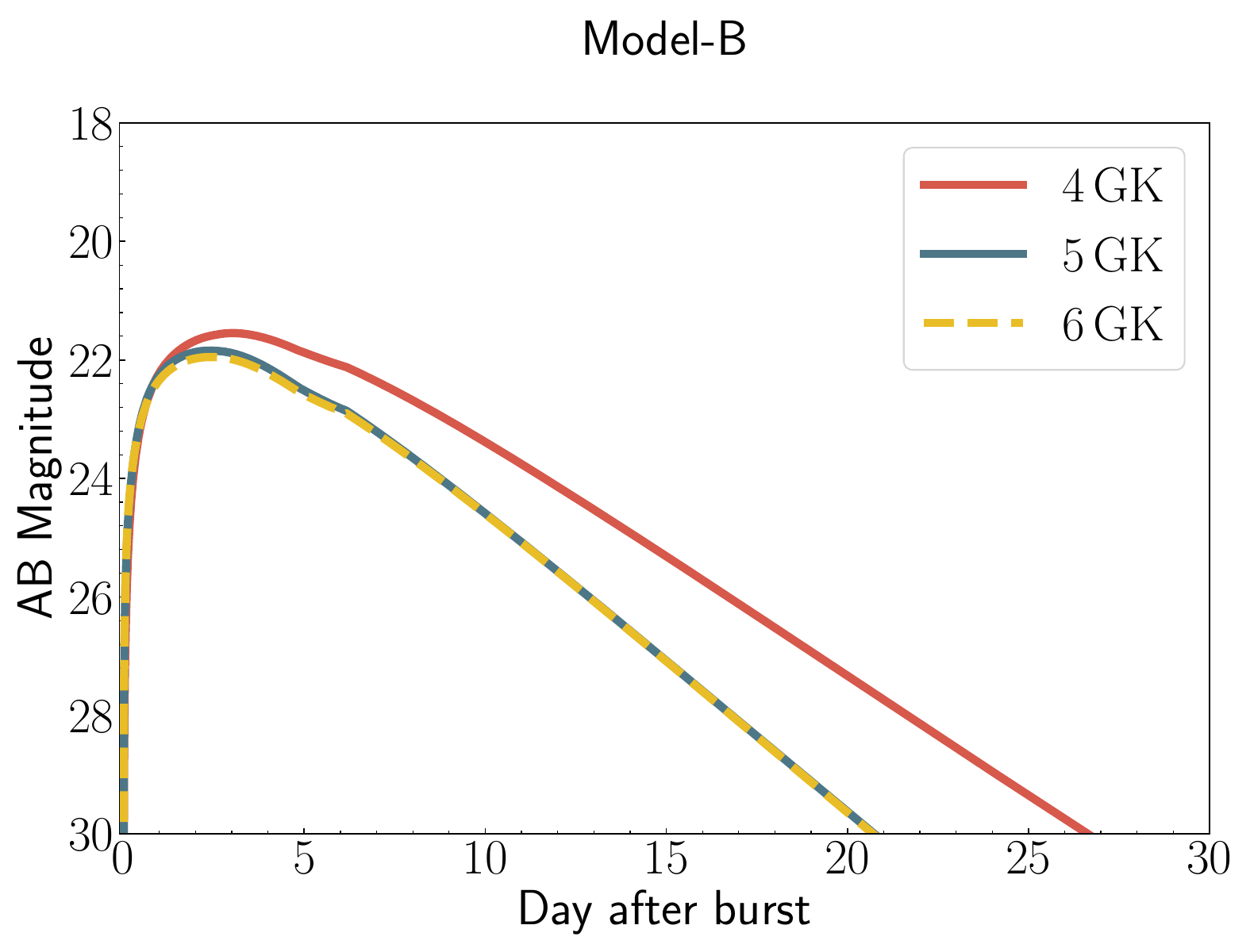}
   \hfill
   \includegraphics[height = 2.55in]{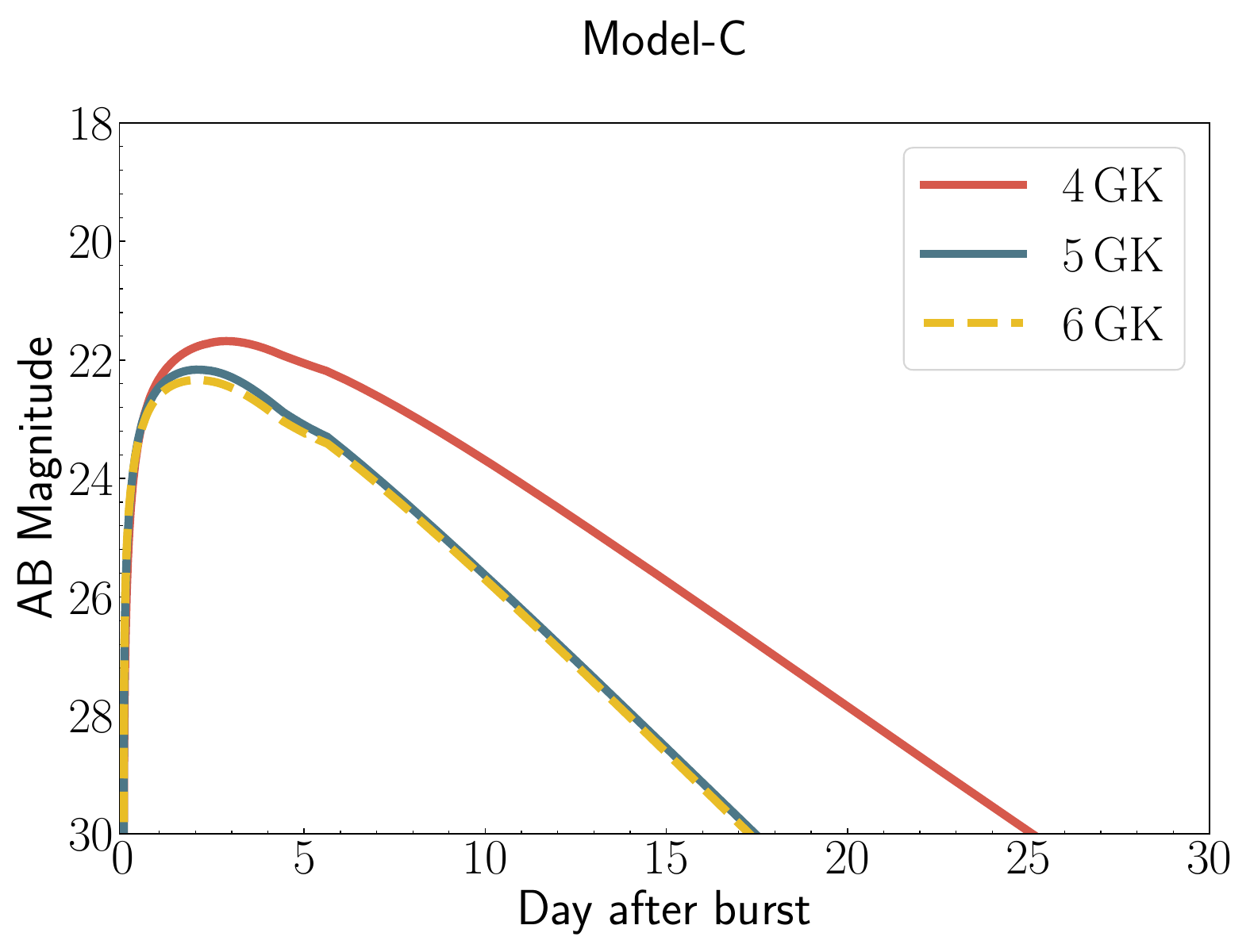}
    \caption{Kilonova-like light curves of the r-band for  Model-A (top panel), Model-B (middle panel), and Model-C (bottom panel) with three different temperatures at $D_{\rm L}=346\rm~Mpc$, respectively.}
    \label{FIG:kilonova}
\end{figure}

\subsection{Numerical Calculation of Light Curve of kilonova-like event}\label{sec:ResultKN}
As mentioned in Section \ref{sec:ResultNuc}, the decay chain of $\rm^{56}Ni\rightarrow{}^{56}Co\rightarrow{}^{56}Fe$ can release a large amount of energy, which is used to heat the ejecta to power the kilonova-like emission. In order to test how bright the kilonova-like emission powered by the radioactive decay can be, following the methods presented in Section \ref{sec:MethodKNe} and \ref{sec:MethodQdot}, we calculate the light curves of kilonova-like emission for the three models with three different temperatures, respectively. Figure \ref{FIG:kilonova} shows the $r-$band light curves of the kilonova-like emission for all three models\footnote{Here, we adopt the luminosity distance $D_{\rm L}=346\rm~Mpc$ which is the distance of GRB 211211A \citep{2022Natur.612..228T}}. For the same given temperature, it is found that the kilonova-like emission of Model-B is the brightest. That is because the peak luminosity is strongly dependent on the mass of the ejecta; namely, the larger the ejecta mass, the higher the peak luminosity of the kilonova-like emission.

On the other hand, by comparing the results for different temperatures within each model, we find that the kilonova-like emission at $T=4$ GK is brighter than that at $T=5$ GK and $T=6$ GK for each model. The reason may be that the abundance of $\rm ^{56}Ni$ at $T=4$ GK is higher than that at $T=5$ GK and $T=6$ GK during the long evolution of the nuclear reaction network, and $\rm ^{56}Ni$ will be decomposed into $\alpha$ particles at $T=5$ GK and above, resulting in a lower abundance of $\rm ^{56}Ni$. However, the brightness of the kilonova-like emission at $T=5$ GK is slightly higher than at $T=6$ GK (see Figure \ref{FIG:kilonova}), indicating that the abundance of $\rm ^{56}Ni$ does not differ significantly between these two temperatures.

\section{Application to the possible kilonova emission associated with GRB 211211A}\label{sec:211211A}
In recent years, the peculiar and nearby long-duration GRB 211211A with a redshift of $z=0.076$ has attracted great attention \citep{2022Natur.612..223R, 2022Natur.612..228T, 2022Natur.612..232Y, 2023ApJ...943..146C, 2023NatAs...7...67G}. The lack of an associated SN with GRB 211211A at such a low redshift, but associated with a possible kilonova emission, together with the long duration of the prompt emission, suggest that its physical origin may be a compact binary star merger, especially, a merger of a WD and an NS \citep{2022Natur.612..232Y, 2023ApJ...947L..21Z}. In Section \ref{sec:result}, we briefly discussed the heavy elements nucleosynthesis and kilonova-like emission from NS-WD mergers. In this Section, from the perspective of heavy elements nucleosynthesis, we will discuss whether the NS-WD merger can indeed power a kilonova-like emission associated with GRB 211211A.

\begin{figure}[!tpb]
    \centering
    \includegraphics[height = 2.55in]{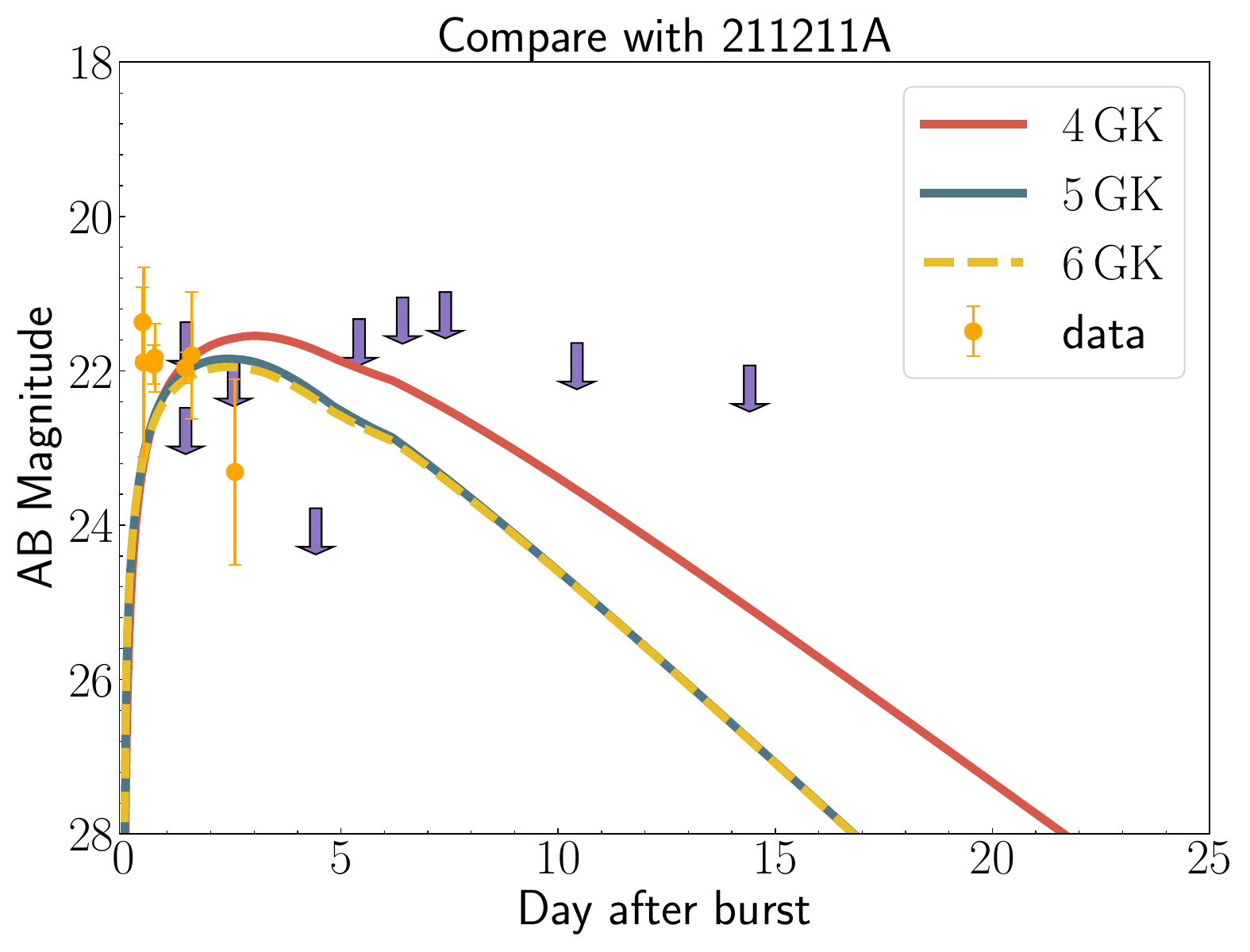}
    \caption{Comparison of calculated kilonova-like emission in r-band for Model-B with three different temperatures at $\rm D_L=346Mpc$ and the observations of the possible kilonova associated with GRB 211211A. The data (orange points) and upper limits (arrows) of the afterglow-subtracted observations are taken from \citep{2022Natur.612..232Y}.}
    \label{FIG:compare211211A}
\end{figure}

In order to compare the calculated kilonova-like emission with that of afterglow-subtracted observations of possible kilonova emission in GRB 211211A, we adopt Model-B, which is the brightest model for powering kilonova-like emission. By collecting the data of possible kilonova emission (or even upper limits) with afterglow-subtracted observation of GRB 211211A from \cite{2022Natur.612..232Y}, we plot the magnitude of possible kilonova emission in r-band for GRB 211211A to compare with the calculated kilonova-like emission from Model-B (see Figure \ref{FIG:compare211211A}). The magnitude of kilonova-like emission at the temperature of $T=4$ GK reaches its peak at about 3 days, and that of $T=5$ GK and $T=6$ GK peaks at about 2.5 days. We find that the observations of the possible kilonova emission of GRB 211211A are lower than those of our calculated model after 1 day. It means that the calculated kilonova-like emission from radioactive decay of the nuclei is sufficient to provide the energy to power the possible kilonova emission associated with GRB 211211A after 1 day of the burst. However, the calculated luminosity from the theoretical model cannot reach such a high luminosity at the early stage, and the afterglow-subtracted observations of the possible kilonova emission associated with GRB 211211A are brighter. It suggests that a single energy source (only radioactive decay) could not provide enough energy to power the possible kilonova emission of GRB 211211A before 1 day after burst. In other words, it may be necessary to add other energy sources, such as the spin-down energy of a newborn magnetar \citep{2022Natur.612..232Y, 2023ApJ...947L..21Z, 2024ApJ...964L...9W}. 

If GRB 211211A is indeed originated from the merger of a WD and an NS, the possible kilonova emission following GRB 211211A would be powered not only by radioactive decay of the nuclei but also by the spin-down energy from the newborn magnetar or other energy sources. If this is the case, the remnant of the WD-NS merger should be a supramassive or stable magnetar. Otherwise, it is difficult to explain the early possible kilonova emission following GRB 211211A by invoking the merger of WD and NS.

Since the brightness of kilonova-like emission depends on the temperature and ejecta mass, the latter is related to the masses of the WD and the NS, the higher mass of the WD and lower mass of the NS in the WD-NS binary system can result in a larger ejecta mass. We calculate the kilonova-like emission again by adopting $T=3$ GK, and find that the brightness of the kilonova-like emission is still fainter than that of $T=4$ GK. Moreover, we also adopt the equal mass of a WD and an NS (e.g., $M_{\rm WD}=M_{\rm NS}=1.25 ~M_{\odot}$) to do the simulation and calculate the kilonova-like emission again and also find that the calculated kilonova-like emission is still fainter than that of observations of possible kilonova associated with GRB 211211A at the early stage.

\section{Conclusion and Discussion}\label{sec:end}
The origin of kilonova-like emission following the LGRBs is currently the subject of much discussion, and mergers of an NS and a WD are also proposed to naturally explain the long duration of GRBs \citep{2022Natur.612..232Y, 2023ApJ...947L..21Z, 2024ApJ...964L...9W}. However, the nucleosynthesis of heavy elements and the brightness of the kilonova-like emission during the mergers of NS and WD are still in debate. In this Letter, by adopting three groups of NSs and WDs with different masses and three different temperatures \citep{2016MNRAS.461.1154M, 2019MNRAS.486.1805Z, 2023ApJ...956...71K}, we studied the heavy elements that are synthesized in NS-WD mergers based on the nucleosynthesis simulations and then compared the results with the observed solid case of possible kilonova emission of GRB 211211A. We find the following interesting results.
\begin{itemize}
\item[1.]  By comparing the results of the nucleosynthesis numerical simulations of three models (i.e., Model-A, Model-B, and Model-C) at different temperatures ($T=4$ GK, 5 GK, and 6 GK), it is found that the final abundance is strongly dependent on both initial temperature and the mass ratio between WD and NS. For the temperature $T=4$ GK, the mass number of all three models peaks at $A = 56$, while the abundance of $\rm ^{4}He$ is substantially higher at $T=5$ GK and $T=6$ GK. Moreover, we also find that the mass number of the produced heavier nuclides is $A_{\rm Model-A} > A_{\rm Model-C} > A_{\rm Model-B}$ at the same temperatures, and the mass number of the produced heaviest nuclides is less than 90 (i.e., $A<90$) with the majority concentrated near the iron peak.  
\item[2.] Kilonova-like light curves are calculated for three models at different temperatures (4, 5, and 6 GK), and it is found that Model-B shows the brightest emission at each given temperature due to its larger ejecta mass. By comparing with adopting different temperatures of each model, the kilonova-like emission at $T=4$ GK is brighter than that of $T=5$ GK and $T=6$ GK, which is likely due to the higher $\rm ^{56}Ni$ production at $T=4$ GK. 
\item[3.] By comparing the calculated kilonova-like emission with that of afterglow-subtracted observations of possible kilonova emission in GRB 211211A, the merger of an NS and a WD cannot be ruled out as the origin of GRB 211211A to produce the possible kilonova emission if the remnant of the WD-NS merger is a supramassive or stable magnetar. Otherwise, it is difficult to explain the early possible kilonova emission following GRB 211211A by invoking the merger of a WD and an NS.
\end{itemize}

We also compare our results with those of other works derived from different models. \cite{2020MNRAS.493.3956Z} adopted 3D hydrodynamical simulations and 2D hydrodynamical-thermonuclear simulations to model the mergers of NSs with CO WDs, which produce faint transient events. In their results, the predicted transient events are much fainter and faster evolving than those of typical Type Ia SNe. For the peak luminosity, the luminosity at $T=4$ GK in our result is not much different from that of their work, with units of magnitude. However, in terms of the time of the peak, the peak time of our results is earlier than that of their work. The reasons for such a difference may be due to different heating rates or different models for calculating the radiative transfer or differences in expansion velocity and ejecta masses. In \cite{2024A&A...681A..41M}, the 3D-magnetohydrodynamics (MHD) simulations of the merger were conducted by using the magnetohydrodynamic moving mesh code AREPO. They claimed that the luminosity of the transients produced by NS-WD mergers could be larger than that of typical kilonovae ($\sim10^{41}\rm~erg\, s^{-1}$), which is also consistent with our results from the aspect of the total luminosity.

In any case, it is difficult to firmly identify the origin of the GRB 211211A-like event only based on its multiwavelength kilonova-like emission. The best way to identify the merger system is to combine with other observational characteristics, such as direct GW detection, or more detailed spectroscopic analysis to look for spectral line features produced by heavier elements, as seen in examples like AT2017gfo, AT2018kzr, and AT2023vfi \citep{2020MNRAS.497..246G, 2023arXiv230800633G, 2025MNRAS.538.1663G,2025ApJ...983L..34L}.

Moreover, GRB 230307A is another interesting event that is similar to that of GRB 211211A \citep{2023arXiv230800633G, 2024ApJ...962L..27D, 2024Natur.626..737L, 2024Natur.626..742Y, 2025NSRev..12E.401S}, and the characteristic of soft X-ray emission suggests that the central engine of GRB 230307A is a magnetar from a binary NS merger \citep{2024ApJ...962L..27D,2024ApJ...963L..26Z,2025NSRev..12E.401S}. \cite{2024ApJ...964L...9W} proposed that the origin of GRB 230307A is also an NS–WD merger, and the late-time kilonova-like emission is powered by spin-down of the magnetar and the radioactive decay of $\rm ^{56}Ni$. More interestingly, the prominent emission feature of GRB 230307A at $\sim2.1\mu m$ detected by James Webb Space Telescope is suggested to be likely attributed to Te \uppercase\expandafter{\romannumeral3} with $A=128$ \citep{2025MNRAS.538.1663G}. Such an element is much heavier than the elements obtained by our simulations from the merger of an NS and a WD. If such an emission line in GRB 230307A is indeed real, it means that at least GRB 230307A originated from the merger of an NS and a WD is very difficult to explain the observed emission line of Te \uppercase\expandafter{\romannumeral3} in the spectrum. In other words, once one detects the emission line of heavier elements ($A>90$) in the spectrum of a GRB 230307A-like event in the feature, the origin of the mergers of an NS and a WD would be ruled out.

Meanwhile, it is worth noting that the system of NS-WD merger is not the only progenitor to explain the type of LGRBs associated with possible kilonova emission. A number of other possible origins are also proposed, such as neutron star-neutron star mergers (NS-NS; \cite{2023ApJ...958L..33G}), NS-BH mergers \citep{2022Natur.612..223R}, WD-BH mergers \citep{1999ApJ...520..650F, 2024MNRAS.535.2800L}, massive accretion disks around black hole \citep{2025ApJ...984...77G}, and low-mass collapsar as the progenitor \citep{2020MNRAS.499.4097Z, 2023PhRvD.108l3038A, 2023ApJ...947...55B, 2024PhRvD.109h3010D}. Further multimessenger observations will be essential to distinguish between these progenitor scenarios and to establish a more comprehensive understanding of the origin of this type of LGRBs.

Our simulation also poses some curious questions. One concern is that the mass number of the heaviest element produced by radioactive decay at $T=5$ GK is larger than that of $T=6$ GK. In general, the reaction rate of nuclides should increase with the rise in temperature. However, the increase of the reaction rate for different nuclides may differ when the temperature rises by 1 GK. Thus, a possible explanation for the above question is that the increase in reaction rates for the nuclides with mass numbers in the range $30<A<60$ is larger than that for those with mass numbers in the range $60<A<80$ when the temperature increases from $T=5$ GK to $T=6$ GK. If this is the case, it can result in more nuclides with mass numbers in the range $30<A<60$, thereby suppressing the production of heavier elements.

\begin{acknowledgements}
We thank Meng-Hua Chen for helpful discussions. This work is supported by the Natural Science Foundation of China (grant Nos. 12494574, 11922301 and 12133003), the Guangxi Science Foundation (grant No. 2023GXNSFDA026007), the Program of Bagui Scholars Program (LHJ), and the Guangxi Talent Program (“Highland of Innovation Talents”).
\end{acknowledgements}

\end{document}